\documentclass[preprint,12pt]{elsarticle}
\usepackage{geometry}
\usepackage[dvipsnames]{xcolor}

\usepackage{amssymb}
\usepackage{amsmath}

\biboptions{sort&compress}

\journal{Extreme Mechanics Letters}

\begin{document}

\begin{frontmatter}



\title{Improving structural 
damage tolerance and fracture energy via bamboo-inspired void patterns}


\author[inst1]{Xiaoheng Zhu\fnref{fn1}}
\author[inst1]{Jiakun Liu\fnref{fn1}}
\author[inst1]{Yucong Hua}
\author[inst1]{Ottman A. Tertuliano\corref{cor1}}
\author[inst1]{Jordan R. Raney\corref{cor1}}

\cortext[cor1]{Corresponding authors}
\fntext[fn1]{These authors contributed equally to this work.}

\affiliation[inst1]{
organization={Department of Mechanical Engineering and Applied Mechanics, University of Pennsylvania}, 
            addressline={220 S 33rd St}, 
            city={Philadelphia},
            postcode={19104}, 
            state={Pennsylvania},
            country={United States} }

\begin{abstract}

Bamboo has a functionally-graded microstructure that endows it 
with a combination of desirable properties, such as high failure strain, 
high toughness, and a low density. As a result, bamboo has been widely
used in load-bearing structures. 
In this work, we study the use of bamboo-inspired void patterns 
to geometrically improve the failure properties of structures made from brittle polymers. 
We perform finite element analysis and experiments on 3D-printed structures to quantify the effect of the shape and spatial distribution of voids on the fracture behavior. 
The introduction of periodic, uniformly distributed voids in notched bend specimens leads to a 15-fold increase in the work of fracture relative to solid specimens. Adding a gradient to the pattern of voids leads to a cumulative 55-fold improvement in the work of fracture. Mechanistically, the individual voids result in crack blunting, which suppresses crack initiation, while neighboring voids redistribute stresses throughout the sample to enable large deformation before failure. 
In addition, we conduct qualitative, low-energy impact experiments on PMMA plates with laser-cut void patterns, illustrating 
the broader potential for this strategy to improve damage tolerance and energy absorption in a wide range of materials systems. 
\end{abstract}






\end{frontmatter}


\begin{figure}[t]
\graphicspath{ {} }
\includegraphics[width=0.6\textwidth]{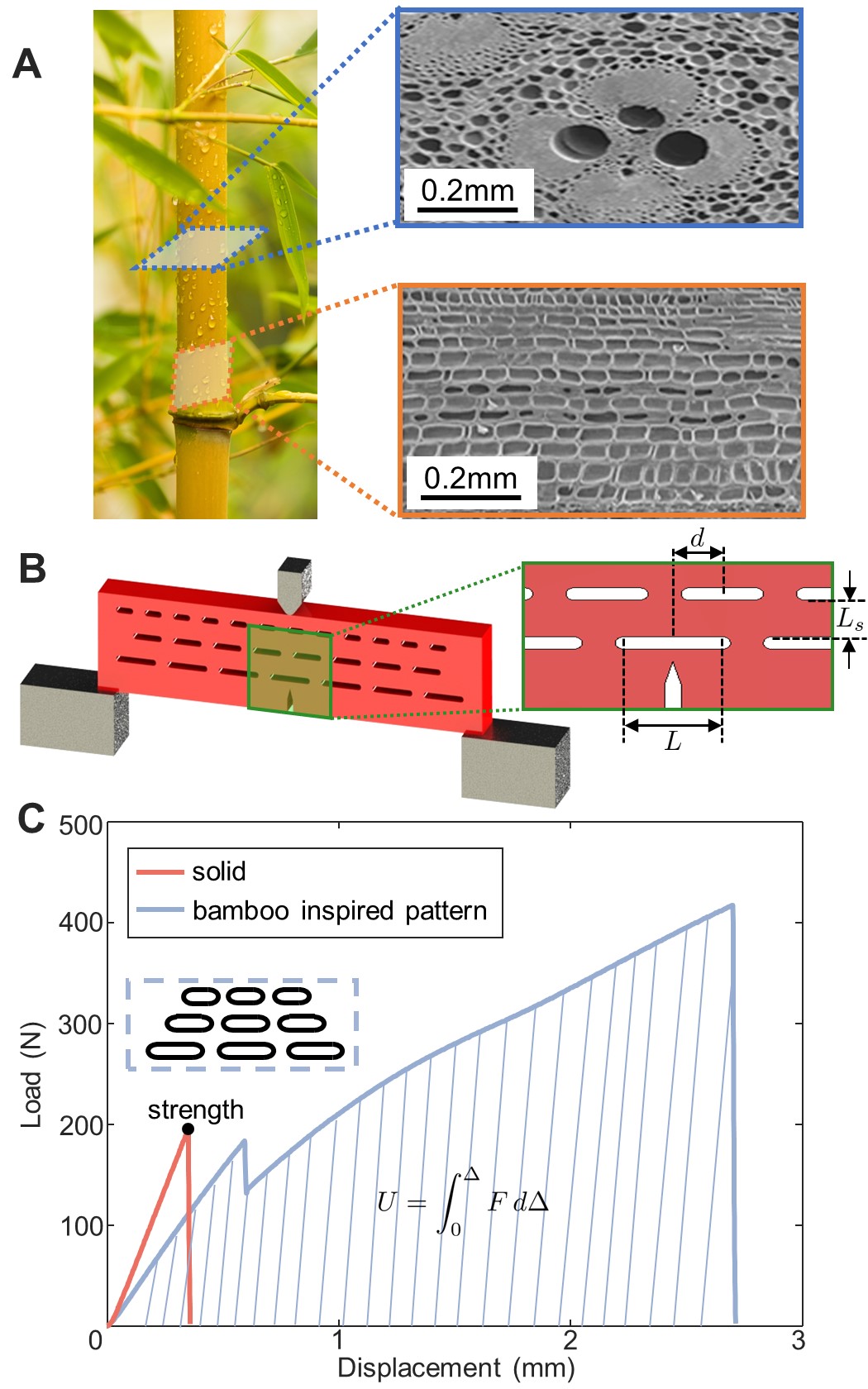}
\centering
    \caption{(A)~Radial and longitudinal sections of bamboo. Images adapted from Ref.~\cite{Tang2019}.
    (B)~Schematic of bamboo-inspired architectures 
    comprising three layers of voids with length $L$, offset $d$, and interlayer distance $L_s$, subjected to three-point bending until failure. 
    (C)~An example experimental measurement of the load-displacement behavior of SENB samples loaded to failure; the energy absorption, strength, and damage tolerance can be significantly improved via the addition of bamboo-inspired void patterns relative to control samples with no voids (solid).}
    \label{fig:intro}
\end{figure}

\section{Introduction}\label{sec:sample1}
Catastrophic brittle failure is encountered in many engineering contexts, ranging from transportation and infrastructure to biomedical devices, bringing with it significant societal costs \cite{Field1971, Berto2014}. 
Catastrophic failure of brittle materials can be attributed to their significant flaw sensitivity. While brittle materials 
can be reasonably resistant to crack initiation, they do not appreciably resist 
crack propagation~\cite{Takano1976}. 
Engineers have few options for avoiding catastrophic failure in these materials, 
such as relying on large safety factors and 
on well-established design principles like increasing 
fillet radii 
to avoid large stress concentration factors. 
Additional strategies have also been developed to improve flaw or crack tolerance, including toughening via 
stress-induced transformations~\cite{Evans1986,Gilbert1997}, 
microcracks in process zones~\cite{Rühle1987}, and grain-localized macro-crack bridging~\cite{Swanson1987,Hutchinson1989}. 
Recent advances in additive manufacturing and multi-material fabrication have enabled additional approaches for 
designing composites with enhanced toughness and impact resistance~\cite{raney2018,mo2022}. Examples include 
beetle-inspired helicoidal composites~\cite{Zaheri2018}, laminated matrix materials~\cite{Bouville2014}, and tough nacre-inspired layered oxide glasses~\cite{Yin2019}. However, these strategies tend to involve materials-specific optimization, since the toughness enhancements strongly depend on the intrinsic interfacial properties of the specific materials used in the composites. 

More recently, researchers in architected materials have begun to explore how 
internal geometric features can be used to improve fracture resistance in lightweight materials~\cite{Jochen2018,Seiler2019,Simon2020,Mo2020,Patel2023}.   
For example, architecting a hexagonal array of holes in aluminum alloy plates can lead to a 50\% reduction in weight while maintaining the same fracture toughness as the solid material~\cite{YLiu2020}. 
The layer-by-layer design freedom of 3D printing has enabled improved material failure properties by controlling micro-structures within a single material~\cite{Mo2020,Pham2019,Lucas2015,mo2023}. These approaches use internal geometry to improve fracture resistance and may be more easily generalizable than methods which rely on unique multimaterial interfaces.
 
Nature provides a number of interesting examples of lightweight materials, such as bone and bamboo, that exhibit both high fracture toughness and strength~\cite{YLiu2020,Ashby2005,Tertuliano2019,Tertuliano2016TheResistance}. 
In all cases, this combination of properties is the result of complex microstructures, including both multimaterial interfaces and geometric features such as voids and internal interfaces. 
For example, the special composition \cite{Liese1986} and gradient-based void patterns seen in bamboo significantly affect its fracture behavior and energy dissipation \cite{Mao2023}. 
The interfacial areas along the fibers 
and the boundaries around parenchyma cells are the preferred route for crack propagation in both radial and longitudinal directions. 
This plays a significant role in determining energy dissipation near the crack tip~\cite{Meisam2015, Habibi2014, Amada2001}.
Beyond bamboo's complex set of composite toughening mechanism, its unique spatial gradient in void geometry can provide insight for developing void-patterning strategies that improve fracture resistance in brittle materials. 
In this work, inspired by the void patterns in the axi-symmetric longitudinal sections of natural bamboo (Fig.~\ref{fig:intro}A), we investigate how gradients in the spatial arrangement and size of voids can lead to higher damage tolerance and energy absorption in brittle materials. 
We find that the work of fracture in brittle polymers can be significantly improved via crack blunting and stress redistribution due to spatially graded voids. Since these results depend merely on the internal geometry of the material system, rather than on intrinsic interfacial properties, this strategy is potentially applicable to a wide range of materials.  


\section{Experiments}
To quantify the effect of different bamboo-inspired void patterns, 
we 3D print single-edge notched bend (SENB) specimens with elongated voids using a linear elastic photopolymer named R11. This polymer has a flexural modulus of 2450~MPa and flexural strength of 75~MPa. This polymer fails in a brittle manner at a fracture strain 0.06. 
The dimensions of the SENB specimens are $67.5$~mm $\times$ $16$~mm $\times$ $6$~mm. 
The edge crack is printed together with the specimen during the printing process. 
The void patterns are printed in front of the crack tip to predictably induce crack growth (Fig.~\ref{fig:intro}B). The bamboo-inspired void pattern comprises three layers of voids with individual void lengths $L$, interlayer void offset $d$, and interlayer distance $L_s$. The voids are rectangles with two semi-circular fillets of radii 0.5~mm at the two ends (see Fig. S1). 
We conducted three-point bend experiments using a commercial quasistatic materials test system (Instron-65SC) using a custom fixture producing three-point loading with a span of 60~mm between supports. The displacement rate was 0.1~mm/s. 

\begin{figure}[htbp]
\includegraphics[width=\textwidth]{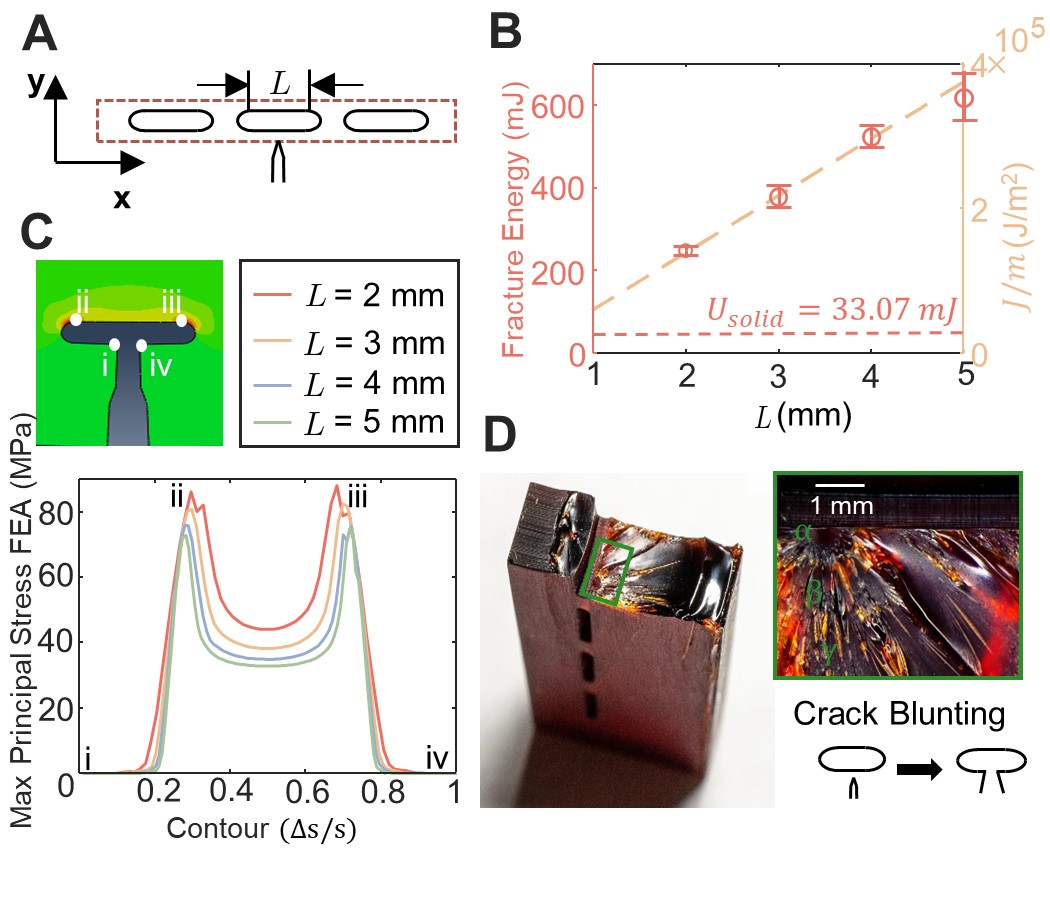}
    \caption{(A) Schematic of 
    a SENB specimen with a single layer of voids of length $L$. (B)~Experimental (red) 
    and analytical (gold) 
    results showing the effect of the length of voids $L$ on the fracture energy of SENB specimens with a single layer of voids; for comparison, the fracture energy of solid SENB samples with no voids is indicated by the horizontal dashed line. 
    (C)~Effect of the length of voids $L$ on the maximum principal stress distribution at the crack tip after the crack propagates into the voids in SENB specimens with a single layer of voids.  
    (D)~An optical image 
    of the crack surface, showing 
    the mirror ($\alpha$), mist ($\beta$), and hackle ($\gamma$) zones.
}
    \label{fig:crack_blunt_std}
\end{figure}
\section{Results}\label{sec:test_and_fea}

The bamboo-inspired voids affect the load-displacement response of the SENB specimens in multiple ways. 
Fig.~\ref{fig:intro}C shows representative load-displacement data for solid and bamboo-inspired specimens. As shown, the addition of the bamboo-inspired void patterns can significantly improve the displacement at failure relative to the solid specimen. There is a characteristic first (small) peak in the load-displacement response of specimens with bamboo-inspired voids, associated with 
the propagation of the crack from the pre-existing notch tip into the nearest void in the first layer, which blunts the crack tip and arrests its propagation. 
In contrast, the solid specimen exhibits unstable crack growth and rapid catastrophic failure, as indicated by the single 
peak in the load-displacement curve. 
Below, 
we systematically investigate the effects of the void size and the spatial arrangement on these failure characteristics.

\subsection{Effect of void length on fracture energy}
To quantify the effect of void geometry on damage tolerance, 
we first consider specimens with only one layer of voids (Fig.~\ref{fig:crack_blunt_std}). 
As in previous work~\cite{Rivlin1953,Luan2022}, 
the fracture energy can be calculated as the area under the load-displacement curve:
\begin{equation}
U =  \int_{0}^{\Delta} F \,d\Delta 
\end{equation}

Crack blunting occurs when the initial crack propagates into the voids. 
As the crack tip enters the void, the effective radius of curvature of the crack tip greatly increases. In this context, crack blunting is quantified by the crack tip opening displacement (CTOD or  $\delta$). For a linear elastic polymer, the relationship between $\delta$ and the energy release rate, taken as the J-integral in the limit of small-scale yielding, is given as:
\begin{equation}
    J =-\frac{\partial U}{\partial A} = m\sigma_{y} \delta
    \label{linear_relation}
\end{equation}
where $m$ is a dimensionless empirical constant that depends on the material strain hardening behavior. The yield strength $\sigma_{y}$ used in this work is 75~MPa. The void induces blunting, and we take the void length as the CTOD, $L$=$\delta$.

Figure~\ref{fig:crack_blunt_std}B shows the effect of void length $L$ (ranging from 1~mm to 3~mm) on the fracture energy, as measured experimentally. In accordance with the energy release rate of Eq.~(\ref{linear_relation}), the fracture energy measured in the experiments is nominally linear with $L$, i.e., $\delta$, suggesting crack blunting by the elongated voids as the main mechanism for increasing fracture energy. 
All samples with voids exhibited larger fracture energy than the solid specimens (i.e., specimens with no voids). 

Specifically, the bamboo-inspired elongated voids distribute stress over a larger area, which can reduce the stress concentration at the two fillet points. 
To quantify this effect further, we used finite element analysis (FEA) to compute the maximum principal stress around the crack tip, a predictor of crack extension direction immediately before initiation  
\cite{Erdogan1963, Maiti1983}. 
To mimic crack blunting in numerical models, we pre-cut the crack from the initial crack tip to the first layer of voids (see Fig.~\ref{fig:crack_blunt_std}C). Then, we applied three-point loading in the model and extracted the maximum principal stresses around the void. Figure~\ref{fig:crack_blunt_std}C shows the maximum principal stress contour 
(around the void) with different void lengths $L$. The stress contour is symmetric about the center line of the middle void because of the symmetry of loading and boundary conditions. As expected, the local maximum of the maximum principal stress occurs at the fillet point of the void. The maximum value of the stress contour is 
inversely 
related to the length of voids; samples with larger voids have a lower maximum principal stress at the crack tip. This stress analysis is consistent with the fracture energy measurements, reinforcing the described crack blunting. 

Optical images of a representative crack surface are shown in Fig.~\ref{fig:crack_blunt_std}D. The fracture surface contains  the mirror ($\alpha$), mist ($\beta$), and hackle ($\gamma$) zones, with increasing roughness. 
The roughness of the fracture surface can be treated as an indicator of the velocity of the crack propagation \cite{Takahashi1984}. The mirror zone is an indicator of slow crack propagation velocity, whereas the hackle zone corresponds to rapid crack propagation. The mirror zone (denoted as $\alpha$) observed in the image is associated with crack blunting, which significantly reduces the speed of crack propagation. Small amounts of experimental misalignment of the loading head on the sample may cause slight asymmetry of the observed fracture zones with respect to crack propagation direction. The fracture energy measurements still demonstrate systematic differences with respect to varying values of $L$, despite the observed asymmetry (Fig~\ref{fig:crack_blunt_std}D, inset). 

\begin{figure}[htbp]
\graphicspath{ {} }\label{fig:f}
\includegraphics[width=\textwidth]{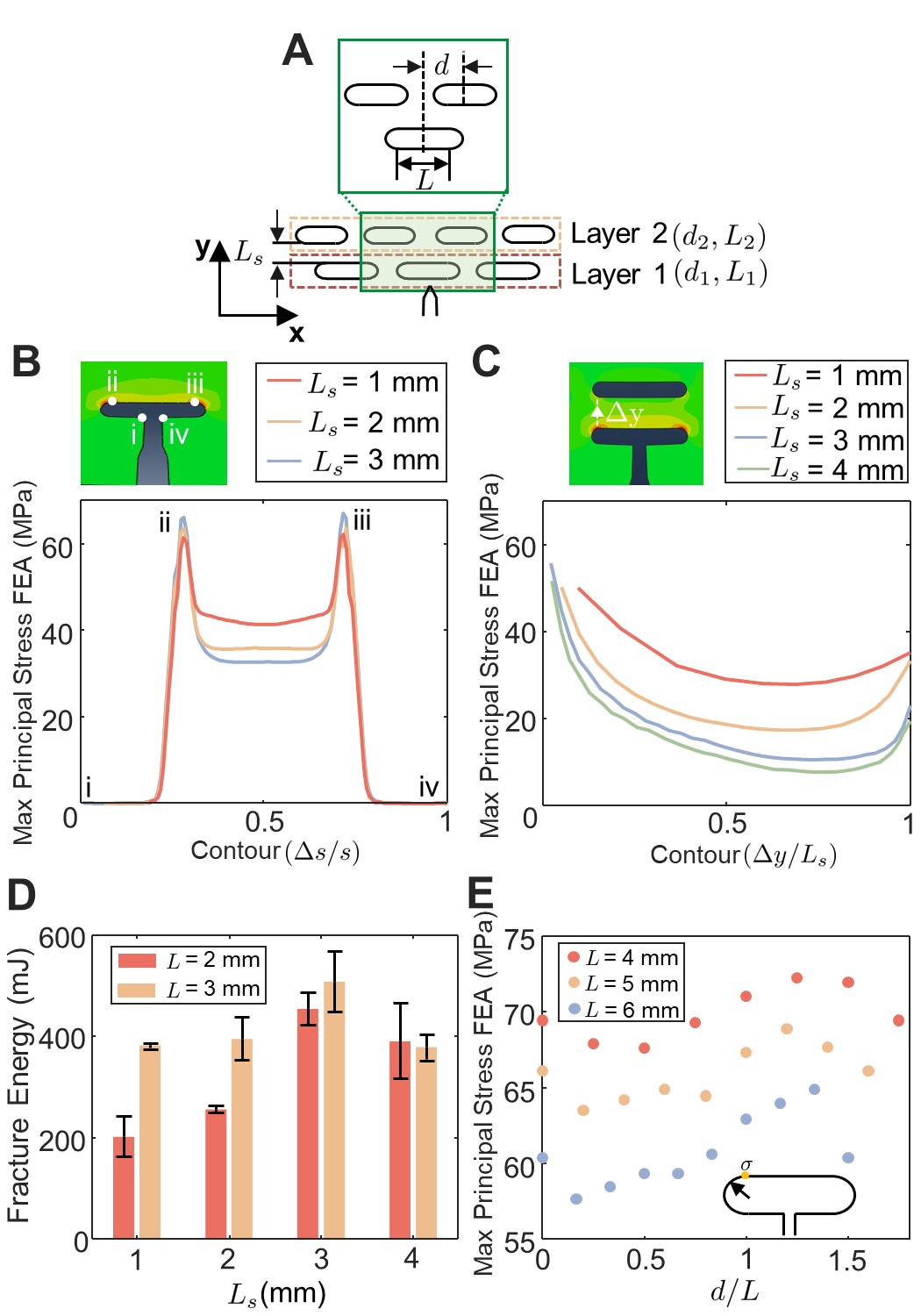}
    \caption{(A)~Schematic of a SENB specimen with two layers of voids of length $L$, offset $d$, and interlayer distance $L_s$. 
    (B)~Effect of the interlayer distance $L_s$ on the distribution of maximum principal stress ahead of the crack tip after the crack propagates into first layer of voids. ($L_1=L_2$, $d_2=0$). 
    (C)~Maximum principal stress distribution from the first layer of voids to the second layer of voids as a function of the interlayer distance $L_s$. 
    (D)~Effect of the interlayer distance $L_s$ on the fracture energy of SENB samples with two layers of voids, as a function of interlayer distance $L_s$ and 
    void length $L$. 
    (E)~FEA of maximum principal stress at the fillet point, after the crack has propagated into the first layer of voids, as a function of 
    the second layer offset $d$ and the void length $L$. 
    }
    \label{fig:crack_interaction_std}
\end{figure}

\subsection{Effect of second layer of voids on crack propagation}


In addition to crack blunting, the interaction of voids between layers can also affect the fracture toughness of micro-architected materials~\cite{YLiu2020,Viggo2002}. It is known that interaction between microcracks affects the stress intensity factor and stress distribution in front of the crack tip, depending on their positions and orientations~\cite{Lam1991}. 

In this section we consider the effect of void interactions in SENB specimens with two layers of voids. As shown in Fig.~\ref{fig:crack_interaction_std}A, the lengths of the voids in the two layers are $L_1$ and $L_2$, respectively, with $L_i$ defining the void length in the $i^{th}$ layer of voids. Moreover the distance between the adjacent layers of voids is defined as $L_s$; the voids between layers may also be offset with respect to one another by a distance $d$. 
We first consider the effect of interlayer distance $L_s$. 
First, numerically we pre-cut the crack (in FEA) from the initial crack tip to the first layer of voids, and extracted the contour plot of the maximum principal stress around the first void in front of the crack tip. Figure ~\ref{fig:crack_interaction_std}B shows the effect of layer distance $L_s$ on the maximum principal stress contour. 
The maximum principal stress at the fillet point slightly increases with increasing layer distance $L_s$. 
To better understand the interactions between voids, we plot the distribution of the maximum principal stress from the fillet edge of the first void to the void in the second layer using FEA.  The maximum principal stress decreased along the path and slightly increased again near the second layer of voids. The maximum principal stress begins to converge for $L_s>3$ mm (Fig.~\ref{fig:crack_interaction_std}C). 
Our simulation results suggest that the layer distance should be larger than 3~mm for this fillet radius to prevent the voids from interacting and to minimize stress concentrations.  
Figure~\ref{fig:crack_interaction_std}D shows the fracture energy for experimental SENB measurements as a function of void lengths $L_1$ and $L_2$ and of interlayer void spacing $L_s$.  
Consistent with the 
numerical results, the fracture energy increases with layer distance $L_s$ and converges at $L_{s}=3$~mm. Considering that we do not increase the overall size of the SENB specimens, decreasing the layer distance reduces the thickness of the material between the two layers, resulting in lower fracture energy. 
However, given the fixed total size of the SENB specimen, when the layer distance is too large, the second layer of voids gets close to the top surface of the specimen, i.e., near the displacement head of the mechanical test system. This alters the boundary conditions and may affect the structural integrity and fracture energy.  

Another parameter we used to control the void interaction effect is the layer offset, $d$. The FEA model predicts the maximum principal stress at the fillet point of the first layer void as a function of the second layer's horizontal offset, as presented in Fig.~\ref{fig:crack_interaction_std}E. The result indicates a periodic relation between maximum principal stress with respect to the layer offset $d$. A second layer of voids with any offset reduces the maximum principal stress when the crack is arrested by the first layer, with respect to samples with no offset. This result indicates the effectiveness of reducing the maximum principal stress at the crack tip through controlling layer offset $d$.

\begin{figure}[htbp]
\graphicspath{ {} }
\includegraphics[width=\textwidth]{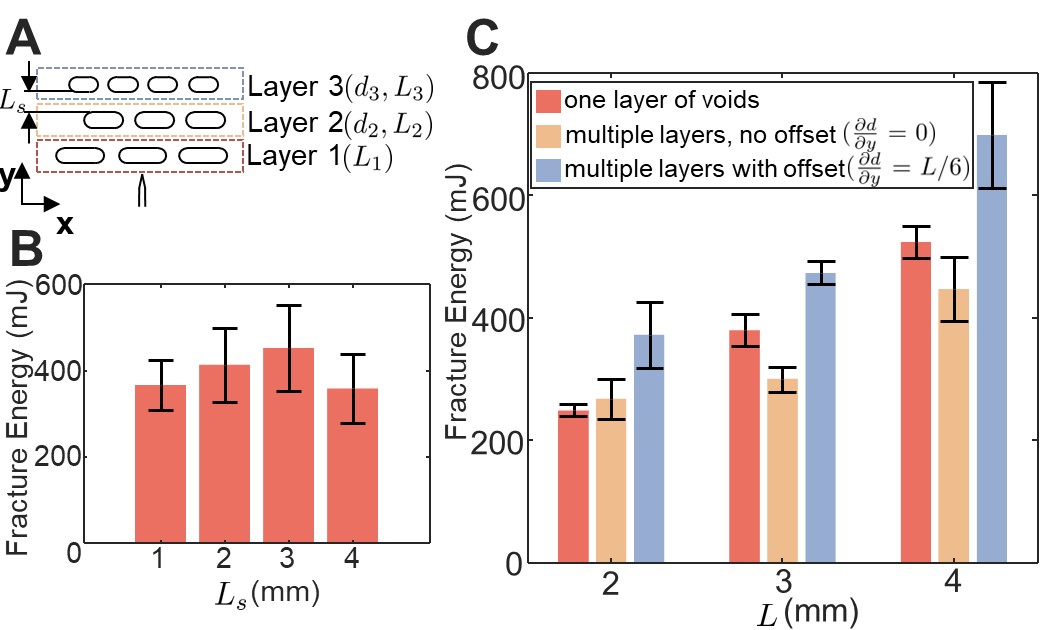}
    \caption{ (A) 
    Schematic of a SENB specimen with three layers of voids, with length $L$, offset $d$, and interlayer distance $L_s$. (B)~Fracture energy as a function of interlayer distance $L_s$ with void length $L$ = 4 mm. (C)~Fracture energy as a function of the void pattern and the void length $L$, 
    with graded void offset $\frac{\partial d}{\partial y}$.
    }
    \label{fig:multi-layer}
\end{figure}

\subsection{Parametric study of multi-layer systems}

Based on the characterization of crack blunting and void interactions  (Fig.\ref{fig:crack_blunt_std},\ref{fig:crack_interaction_std}), multiple layers of voids can be further used to redistribute stress, delocalize strain, and achieve higher fracture energy. Natural materials, like bamboo, are generally not uniform; they use heterogeneity and gradients to improve 
mechanical and physical properties under the loading scenarios critical to their survival. 
We performed a parametric study to explore the effect of 
graded, bamboo-inspired void patterns on the fracture energy of SENB specimens~(Fig.\ref{fig:multi-layer}A). 
To mitigate the negative void interaction effect brought by layers of voids, we first found appropriate interlayer void spacing $L_s$ for the multi-layer system without gradients. Then we kept the $L_s$ constant for the remaining study of bamboo-inspired graded void patterns. 
Figure~\ref{fig:multi-layer}B shows the fracture energy of a multi-layer system with uniform void sizes as a function of layer distance. We chose void size $L$ = 4 mm to maximize crack blunting. Similar to the trend for two-layer systems, the peak fracture energy is achieved with interlayer distance $L_s=3$~mm. 
We then fixed the interlayer distance to $L_s=3$~mm for the remainder of the study. After found appropriate interalyer distance, we parameterized graded, bamboo-inspired void patterns.
With the given void length in the first layer, the parameters for the second and third layers were determined by defining a gradient along the interlayer spacing $L_s$ in Eq.(\ref{eq:gradients}).
\begin{subequations}
\begin{align}
   \frac{\partial L}{\partial y} &=  \frac{L_{i-1}-L_i}{L_s} \\
   \frac{\partial d}{\partial y} &=  \frac{d_{i-1}-d_i}{L_s}
\end{align}
\label{eq:gradients} 
\end{subequations}

Figure \ref{fig:multi-layer}C shows the effect of layer offset $d$ on the fracture energy of multi-layer systems. Adding layers of voids with offsets ($\frac{\partial d}{\partial y}$= 0) to SENB specimens with a single layer of voids produces very little improvement in the fracture energy, and may actually decrease it. 
In contrast, multiple layers with offset defined by  $\frac{\partial d}{\partial y}$= $L/6$~mm can improve the fracture energy with various void lengths. Laterally offsetting voids from layer to layer reduces the maximum principal stress at the crack tip after the crack propagates into the first layer of voids (as shown in Fig.~\ref{fig:crack_interaction_std}E). This offset affects the second peak in the load-displacement curve, and, therefore, the fracture energy. 

\begin{figure} [htbp]
\graphicspath{ {} }
\includegraphics[width=\textwidth]{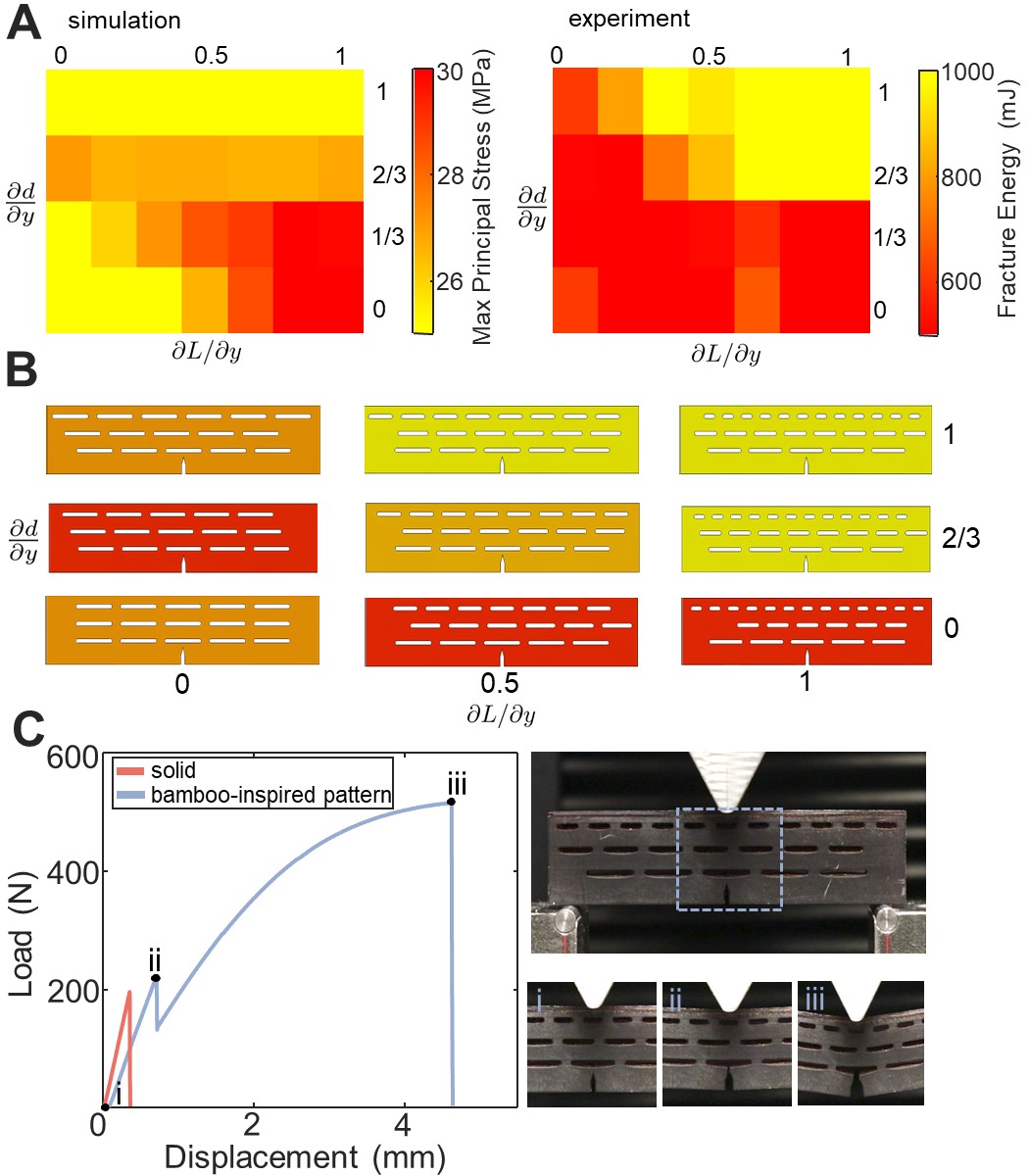}
    \caption{ (A) Parametric study of the effect of gradient void patterns on the maximum principal stress at the fillet point and the fracture energy. 
    (B) Schematic illustrating the versatility of SENB specimens with gradient voids. (The color of the samples corresponds with their fracture energy in experiments.)
    (C)~(left) Fracture behavior of a bamboo-inspired SENB sample relative to a solid SENB sample; (right)~optical images of the experiment, with labels \emph{i, ii,} and \emph{iii} corresponding to the indicated load-displacement points. }
    \label{fig:paramtric study}
\end{figure}

After noticed the improvement brought by void patterns with offset, we did the parametric study on graded, bamboo inspired void pattern using both experimental and numerical methods (see Fig.~\ref{fig:paramtric study}A). We chose the void length in the first layer to be $L_1=8$~mm, resulting in five periodically spaced voids in each layer of voids. Given the fixed dimensions of the SENB specimens, this choice balances the desire for a larger number of voids per layer with the need for a void size that is sufficiently large so as to avoid manufacturing errors. Then, we set the difference in $L$ and $d$ from one layer to the next to have an upper bound equal to the interlayer spacing, $L_s$, i.e., the gradients in Eq.(\ref{eq:gradients}) are defined by ($\frac{\partial L}{\partial y}$) and ($\frac{\partial d}{\partial y}$) $\in$ [0,1]. 
The simulations suggest that for SENB samples with multiple layers of voids, adding layer offset can decrease the maximum principal stress at the fillet point, except for the uniform case ($\frac{\partial L}{\partial y}$=$\frac{\partial d}{\partial y}$ = 0 ). Our experiments show that larger gradients in void size and offset result in larger fracture energies (see Fig.\ref{fig:paramtric study}A). 
We note that the maximum principal stresses computed from the simulations can only provide insight about the trends in fracture energy observed experimentally. The symmetry in the simulation suggests equal probability of crack propagation at the void edges. However, any amount of misalignment in the position of the applied load or imperfections in the printing process would cause crack propagation at one void edge rather than both. 

The bamboo-inspired void patterns in SENB specimens cause a significant improvement in fracture energy relative to the SENB specimens without any voids (Fig.~\ref{fig:paramtric study}C). The introduction of periodic void arrays in notched specimens can lead to a 15-fold increase in the fracture energy relative to solid SENB specimens. 
Adding a gradient to the void patterns can additionally increase the fracture energy, resulting in a 55-fold increase over solid samples. 
Figure~\ref{fig:paramtric study}C shows a comparison of loading curves of the solid SENB specimen and the SENB specimen which achieved the highest fracture energy in the parametric study. The 
latter is composed of three layers of voids with a large gradient in void length and large layer offsets, $\frac{\partial L}{\partial y}$ and $\frac{\partial d}{\partial y}$, respectively. The long first layer of voids can effectively magnify the crack blunting effect. The second and the third layers of voids can reduce the maximum principal stress at the crack tip and improve the fracture energy. 
In three-point bending experiments, the specimen demonstrated remarkable fracture strain (maximum displacement), which contributes to the improvement in fracture energy. 

\begin{figure} [htbp]
\graphicspath{ {} }
\includegraphics[width=\textwidth]{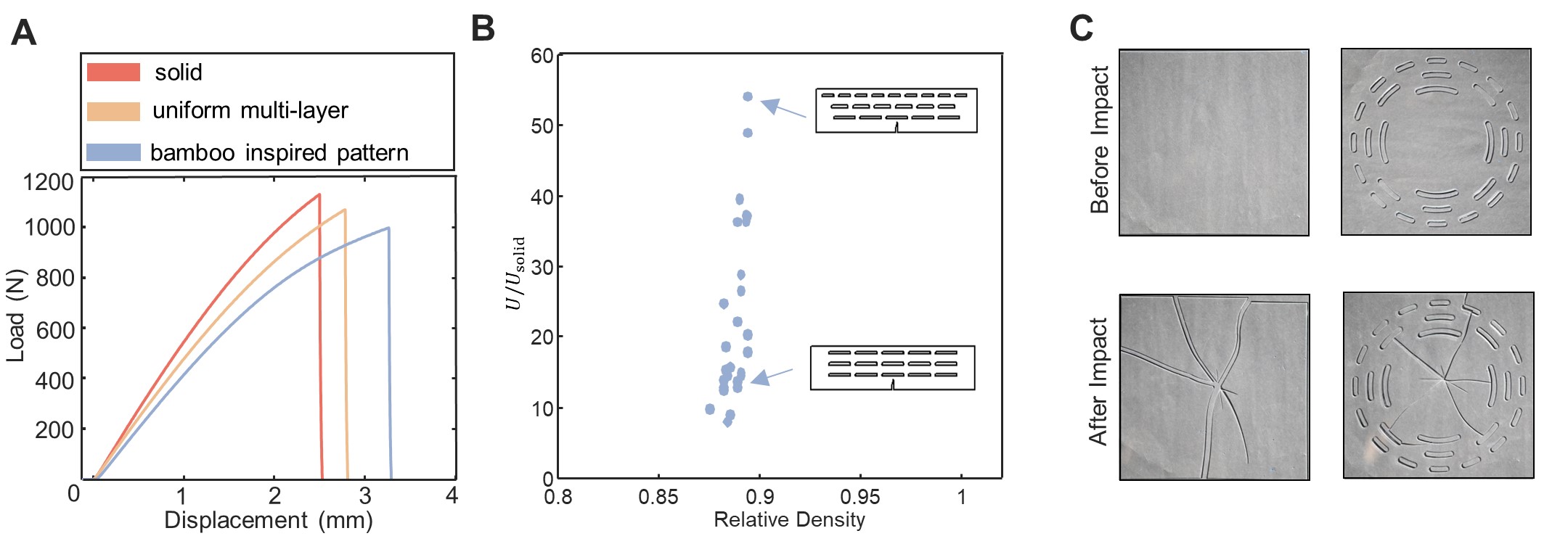}
    \caption{(A) Experimental measurements of the fracture behavior of unnotched bending specimens; specimens consist of solid material, uniform layers of voids, or bamboo-inspired void patterns, as indicated by the legend. 
    (B)~Relative fracture energy vs. relative density for bamboo-inspired structures, based on experimental measurements. 
    (C)~Optical images of PMMA plates subjected to low-energy impact experiments; images show the behavior of plates consisting of solid material (left) or bamboo-inspired void patterns (right).}
    \label{fig:application}
\end{figure}

\section{Discussion and concluding remarks}

While this work has focused on the effect of bamboo-inspired void patterns on the energy associated with failure, given the well-known trade-off between toughness and strength it is worth briefly discussing the effect of these void patterns on strength. 
In marked contrast with notched specimens, the force-displacement behavior of unnotched samples exhibits only one peak (compare Fig.~\ref{fig:paramtric study}C with Fig.~\ref{fig:application}A). Unnotched specimens with bamboo-inspired, gradient voids show $20\%$ less strength than solid specimens, despite the large improvement in fracture energy measured for the notched specimens. 
This is due to the fact that the strength of the brittle material itself controls the fracture of the unnotched samples (and any voids will concentrate stress). In contrast, the crack tolerance of the structure/material controls the work of fracture of the notched samples. The bamboo-inspired void patterns improve the structural compliance. As shown in Fig.~\ref{fig:application}B, which plots 
fracture energy vs. relative density, bamboo-inspired void patterns cause order-of-magnitude improvement in fracture energy despite a roughly $10\%$ decrease in relative density. 

To better understand why these architectures achieve such a large increase in work of fracture (at the cost of a small decrease in strength), we take the case of samples with two layers of voids as an example: the stress distribution ahead of the crack tip is not monotonic (see Fig.~\ref{fig:crack_interaction_std}C). Starting from the crack tip (the fillet point), the stress decreases first and increases again for the region near the second layer of voids. In other words, multiple layers of voids act as multiple stress concentration points that redistribute the stress distribution of the sample. Similarly, Fig.~\ref{fig:paramtric study}Ciii shows the same effect in samples with three layers of voids: the second and third layers of voids cause the deformation to distribute throughout the sample during loading, i.e., load sharing~\cite{Simon2020}. The voids 
increase structural compliance, which increases the strain energy prior to failure. 

As a conceptual demonstration of the potential utility of these architectures, we produced bamboo-inspired void patterns in PMMA plates via a laser cutter. These were subjected to low-energy impact experiments. Figure~\ref{fig:application}C shows optical images before and after impact for both solid plates (left) and the plates with bamboo-inspired void patterns (right). The solid plates catastrophically fractured into multiple pieces. In contrast, the plates with bamboo-inspired void patterns trapped the cracks as they extended radially from the impact site, thereby preventing the plate from breaking into separate pieces. 

As mentioned previously, natural bamboo consists of densified vascular bundles and parenchyma cells, resulting in constitutive behaviors far more complex than what we have explored here. In this study, we limited our focus to geometric effects in brittle materials. The ductility of materials would certainly affect the failure characteristics of structures with architected void patterns. Combining materials-intrinsic ductility with the geometrically-enabled structural compliance may be a promising direction for future research. 

In summary, the introduction of periodic, uniformly distributed voids in notched bend specimens leads to a 15-fold increase in the fracture energy relative to notched specimens with no voids. Adding a gradient to the pattern of voids can lead to a cumulative 55-fold increase in fracture energy relative to the solid specimen. Furthermore, this order-of-magnitude improvement in the fracture energy results in only a 20$\%$ reduction of strength. 
This work demonstrates the potential of using the internal architecture to improve the failure performance of 
brittle materials. 
These principles can be applied to other brittle material systems which are susceptible to catastrophic failure, and could lead to important practical improvements in applications such as 
infrastructure, semiconductors, and protective gear.

\section{Materials and Methods}

\subsection{Fabrication of samples}
The structures were printed with a commercial digital light projection 3D printer (EnvisionT
EC Vida HD). The printer has a build volume of $96 \times 54 \times 100 \, mm$ with XY resolution of 25~$\mu$m and a z step size of $25 \, \mu m$. Due to the size limit of the 3D printer, the dimensions of SENB specimens tested in this study were 67.5 mm $\times$ 16 mm $\times$ 6 mm (length $\times$ width $\times$ thickness). The aspect ratio of the loading span to the length of the sample is 1.25. The edge crack was printed together with the specimen during the printing process. The printer takes grayscale images (1920x1080) as inputs and projects these into the resin. The material used in this work is a photopolymer named R11, a brittle material. The mechanical properties of R11 were measured via uni-axial tensile tests at a strain rate of 0.002/s using an Instron-65SC.

\subsection{Fracture tests}
Fracture testing was performed by conducting three-point bend tests under displacement control with a loading rate of 1~mm/s 
using an Instron-65SC at room temperature. The span between two supporting points was 60~mm. At least three specimens were tested for each type of experiment. 

\subsection{Finite element analysis}
Finite element analysis for static loading of linear elastic materials was carried out using Abaqus. The specimen geometries were generated in Python and SOLIDWORKS. The geometry was modeled using four-noded, plane strain elements. Example meshes for the case are shown in supplementary for reference.

\bibliographystyle{elsarticle-num-names} 
\bibliography{bamboo-paper-1-refs}





\end{document}